\documentclass[twocolumn,preprintnumbers,floats,prd,amssymb,floatfix,nofootinbib,balancelastpage,superscriptaddress,amsmath]{revtex4-1}

\usepackage[utf8]{inputenc}
\usepackage{CJKutf8}
\usepackage{amssymb}
\usepackage{amsmath}
\usepackage{amsfonts}
\usepackage{graphicx}
\usepackage{color}
\usepackage{xspace}
\usepackage{comment}
\usepackage{hyperref}
\usepackage[normalem]{ulem}
\usepackage[section]{placeins}
\usepackage{afterpage}
\usepackage{slashed}
\usepackage{enumerate}
\usepackage{enumitem}
\usepackage{caption}
\usepackage{subfigure}
\usepackage{float}
\usepackage{ulem}
\usepackage{verbatim}
\usepackage{multirow,rotating}
\usepackage[dvipsnames]{xcolor}
\usepackage{braket}

\definecolor{dcolour}{rgb}{.5, .5, .5}
\def\gsim{\raise0.3ex\hbox{$\;>$\kern-0.75em\raise-1.1ex\hbox{$\sim\;$}}}
\def\lsim{\raise0.3ex\hbox{$\;<$\kern-0.75em\raise-1.1ex\hbox{$\sim\;$}}}
\def\gsim{\raise0.3ex\hbox{$\;>$\kern-0.75em\raise-1.1ex\hbox{$\sim\;$}}}
\def\lsim{\raise0.3ex\hbox{$\;<$\kern-0.75em\raise-1.1ex\hbox{$\sim\;$}}}

\newcommand{\ba}[1]{\begin{eqnarray} \label{(#1)}}
	\newcommand{\ea}{\end{eqnarray}}

\newcommand{\TAB}[1]{Table~\ref{#1}}

\newcommand{\GeV}{\,{\mathrm{GeV}}}

\newcommand{\la}{\langle}
\newcommand{\ra}{\rangle}

\begin{document}
\captionsetup[figure]{justification=raggedright,singlelinecheck=false}
\title{Theoretical study of the $B^+\to D^-D_s^{+}\pi^+$ reaction }
\author{Xuan Luo}
\email{xuanluo@ahu.edu.cn}
\affiliation{School of Physics and Optoelectronics Engineering, Anhui University, \\
		Hefei, Anhui 230601, People’s Republic of China}
\author{Ruitian Li}
\affiliation{Institute of Theoretical Physics, School of Physics, Dalian University of Technology, \\ 
	No.2 Linggong Road, Dalian, Liaoning, 116024, People’s Republic of China}
\author{Hao Sun}
\email{haosun@dlut.edu.cn}
\affiliation{Institute of Theoretical Physics, School of Physics, Dalian University of Technology, \\ 
	No.2 Linggong Road, Dalian, Liaoning, 116024, People’s Republic of China}

\begin{abstract}
Prompted by the recent discoveries of $T_{c\bar{s}0}(2900)^{++}$ in the $D_s^+\pi^+$ invariant mass distribution of $B^+\to D^-D_s^+\pi^+$ process, we present a model that hopes to help us investigate the nature of $T_{c\bar{s}0}(2900)^{++}$ by reproducing the mass  distribution of $D^-\pi^+, D_s^+\pi^+$ and $D^-D_s^+$ in $B^+ \to D^-D_s^+\pi^+$ decays. 
The structure of the triangular singularity peak generated from the $\chi_{c1}D^{*+}K^{*+}$ loop near the $D^{*+}K^{*+}$ threshold is considered in our model may be the experimentally discovered resonance-like state structure $T_{c\bar{s}0}(2900)^{++}$. 
In addition, we employ a coupled-channel approach to describe the dominant contribution of the $D\pi$ $S\text{-wave}$ amplitude, and also consider other  excitations. 
 Our model provides a well fit to the invariant mass distributions of  $D^-\pi^+, D_s^+\pi^+$ and $D^-D_s^+$ simultaneously.
	
\end{abstract}
\vskip10mm
\maketitle
\flushbottom	

\section{Introduction}
\label{I}
Over the last two decades, many resonance states have been discovered experimentally that cannot be explained by the traditional quark model~\cite{10.1093/ptep/ptac097,Guo:2017jvc,Hosaka:2016pey}. These states are called exotic hadron states.
The first to be discovered was the exotic state of $X(3872)$ observed by the Belle Collaboration in 2003~\cite{Belle:2003nnu}, which was later verified by other experiments~\cite{BaBar:2004iez,CDF:2003cab,D0:2004zmu,LHCb:2011zzp}.

 In the year of 2020, the LHCb Collaboration reported the discovery of two new structures $X_0(2900)$ and $X_1(2900)$ in the $D^-K^+$ invariant mass distribution on the $B^+ \to D^+D^-K^+$ process~\cite{LHCb:2020bls,LHCb:2020pxc}. Interestingly, both $X_0(2900)$ and $X_1(2900)$ are all fully open flavor states and their minimal quark components are four different quark $ud\bar{s}\bar{c}$. Therefore, many theoretical works have been dedicated to exploring their properties~\cite{He:2020jna,Karliner:2020vsi,Yang:2021izl,Wang:2021lwy,Dai:2022qwh,Liu:2020orv,Wang:2020xyc,Zhang:2020oze,He:2020btl,Xue:2020vtq,Lu:2020qmp}. 

Recently, the LHCb collaboration studied the $B^0 \to \bar{D}^0D_s^+ \pi^-$ and $B^+ \to D^-D_s^+\pi^+$ processes~\cite{LHCb:2022lzp,LHCb:2022sfr} and observed a new double-charged $\text{spin-}0$ open charm tetraquark candidate and a neutral partner in the $D_s\pi$ decay channel with their reported masses and widths: 
\begin{equation}
	\begin{aligned}
		T_{c\bar{s}0}(2900)^{++} :&\quad M= 2.922 \pm 0.014(\text{GeV})\\
		&\quad \Gamma= 0.161 \pm 0.033(\text{GeV}),
	\end{aligned}
\end{equation}
and
\begin{equation}
	\begin{aligned}
		T_{c\bar{s}0}(2900)^0:& \quad M= 2.871 \pm 0.012 (\text{GeV}) \\
		&\quad \Gamma= 0.135 \pm 0.025 (\text{GeV}),
	\end{aligned}
\end{equation}
respectively.

 From the process observed, the minimal quark components of the  $T_{c\bar{s}0}(2900)^{++}$ and $T_{c\bar{s}0}(2900)^0$ are $c\bar{s}\bar{d}u$ and $c\bar{s}\bar{u}d$, respectively~\cite{LHCb:2022lzp,LHCb:2022sfr}. 
 The appearance of the exotic state has aroused the interest of numerous theoretical physicists, because the former particle is the first observed doubly charged tetraquark state. 
 A number of theories have been proposed to  provide further guidance to explore the nature of $T_{c\bar{s}0}(2900)$.
  Theoretical works have investigated $T_{c\bar{s}}(2900)$ properties using the QCD sum rule approach~\cite{Yang:2023evp,Lian:2023cgs,Liu:2022hbk}, suggesting that it could be a scalar tetraquark state of $c\bar{s}q\bar{q}$. And there is also a study that has investigated the tetraquark state properties of $T_{cs}$ and $T_{c\bar{s}}$ under coupled-channel calculation based on the constituent-quark-model~\cite{Ortega:2023azl}.

The authors in Ref.~\cite{Molina:2022jcd} proposed that $T_{c\bar{s}0}$ can be interpreted as a threshold effect resulting from the interaction between the $D^*K^*$ and $D_s^*\rho$ channels. 
It has been suggested that $T_{c\bar{s}0}$ could be a dynamical effect arising from the triangular singularity in Ref.~\cite{Ge:2022dsp}.
Within the framework of the local hidden gauge approach, the authors studied the coupled-channel $D_s^*\rho - D^*K^*$ interactions~\cite{Duan:2023lcj}.	
Furthermore, the observed mass of $T_{c\bar{s}0}$ is closely aligned with the $D^*K^*$ threshold, suggesting that it has the potential to be a promising candidate for a molecular state composed of $D^*$ and $K^*$.
Then there were also numerous works investigating the nature of the molecular state of $T_{c\bar{s}0}$~\cite{Chen:2022svh,Ke:2022ocs,Yue:2022mnf,Agaev:2022eyk,Duan:2023qsg,Yue:2023qgx}.
In Ref.~\cite{Agaev:2022duz}, the authors also analyzed the interpretation of $T_{c\bar{s}0}$ state as $D_s\rho$ hadronic molecules using the QCD sum rule method.

Until now, the nature of $T_{c\bar{s}0}(2900)^{++}$ has yet to be definitively determined and remains subject to ongoing investigations and research. Therefore, we hope that fitting the three invariant mass distributions from the experimental results will help us to determine the characteristics of $T_{c\bar{s}0}(2900)^{++}$.

It is significant to emphasize that the $D\pi$ $S\text{-wave}$ significantly affected the $D\pi$ invariant mass distribution in the $B^+\to D^-D_s^+\pi^+$ process. In the experiment, the distribution of $D\pi$ $S\text{-wave}$ was improved by introducing a $0^+$ quasi-model-independent~\cite{LHCb:2016lxy} description. 

Thus, in this work, we have developed a model with few parameters that can be fitted simultaneously to the three invariant mass distributions  $M_{D^-\pi^+},$ $M_{D_s^+\pi^+}$ and $M_{D^-D_s^+}$ in $B^+ \to D^-D_s^+\pi^+$ process.  We have taken into account Breit-Wigner (BW) amplitudes, a triangle loop amplitude associated with $T_{c\bar{s}0}(2900)^{++}$, and a unitary coupled channel effect related to $D\pi$ $S\text{-wave}$ amplitude.
The organization of the paper is as follows.  In Sec.~\ref{II}, we present the theoretical formalism of the process $B^+ \to D^-D_s^+\pi^+$ . The numercial results of the calculations and discussion are presented in the Sec.~\ref{III}. At the end, we make a brief summary  in the Sec.~\ref{IIII}.

\section{framework}
\label{II}

\begin{figure*}[htpb]
	\begin{center}
		\includegraphics[width=0.80\textwidth]{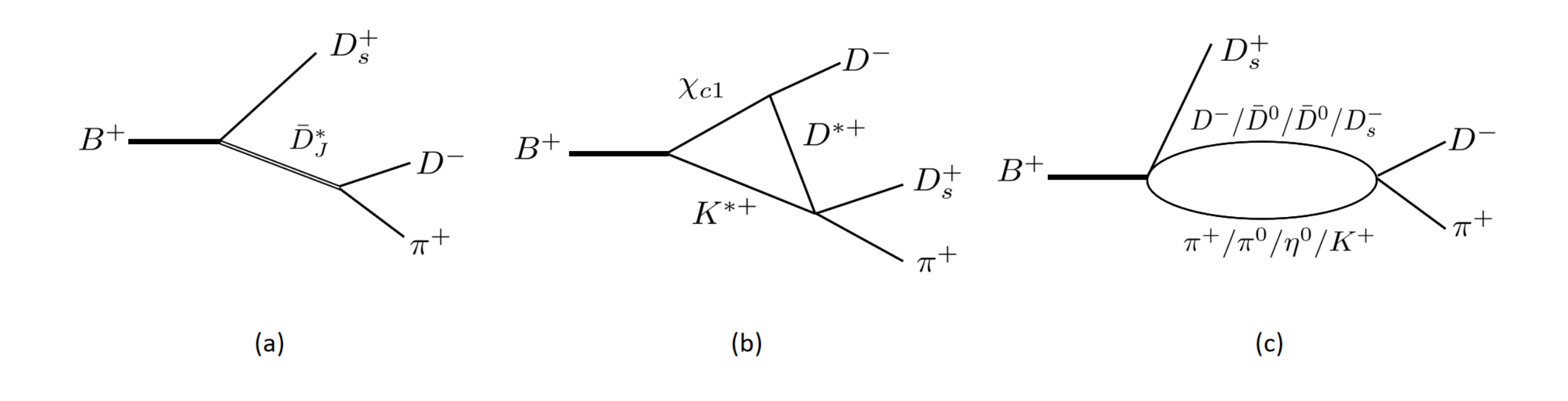}
	\end{center}
	\caption{$B^+ \to D^-D_s^+\pi^+$ considered in this work: (a) $\bar{D}^*_J$ $\left[\bar{D}^*(2007)^0, \bar{D}_2^*(2460), \bar{D}^*_1(2600)\right]$ excitations; (b)  triangle loop; (c) coupled-channel.  }
	\label{fig1}
\end{figure*}

We considered the contributions of the Feynman diagrams depicted in FIG.~\ref{fig1} to process $B^+ \to D_s^+D^-\pi^+$. We derive the relevant amplitudes by writing the effective Lagrangians of the relevant hadrons and their matrix elements, then combining them in accordance with the time-ordered perturbation theory. We considered three kinds of mechanisms: the $\bar{D}^*_J$-excitations of FIG.~\ref{fig1}(a), triangle loop of FIG.~\ref{fig1}(b) and unitary coupled channel effect of $D\pi$ $S\text{-wave}$ of FIG.~\ref{fig1}(c). For convenience, here we use  $\bar{D}^*_J$ to describe this three mesons: $\bar{D}^*(2007)^0$, $\bar{D}^*_1(2600)$ and $\bar{D}^*_2(2460)$.  We label the mass, width, energy, momentum and  polarization vector of a particle $A$ as $m_A$, $\Gamma_A$, $E_A$, $\vec{p}_A$ and $\vec{\epsilon}_A$.  The scattering amplitudes of all the Feynman diagrams that we are thinking about will be discussed below.

\subsection{The mechanism of $\bar{D}^*_{J}$ excitations}
We take the Breit-Wigner forms to describe the $\bar{D}^*_{J}$ excitation mechanisms of FIG.~\ref{fig1}(a). In this work we have taken $\bar{D}^*(2007)^0$, $\bar{D}^*_1(2600)$ and $\bar{D}^*_2(2460)$ into consideration since the data~\cite{LHCb:2022lzp} shows that they have effects on the final states invariant mass distributions in $B^+ \to D_s^+D^-\pi^+$ process.

For $B^+ \to D_s^+ \bar{D}^*(2007)^0 \left[ \bar{D}^*_1(2600)\right]  \left(1^-\right)$ vertex only to be $p\text{-wave}$, and the decay of the $\bar{D}^*(2007)^0 \left[ \bar{D}^*_1(2600)\right]$ to $D^-\pi^+$ also involves $p\text{-wave}$. The expressions for amplitudes $\bar{D}^*(2007)^0$ and $\bar{D}^*_1(2600)$ are:
\begin{equation}
\begin{aligned}
A_{\bar{D}^*(2007)^0} =& c_{\bar{D}^*(2007)^0}\\ & \times \frac{\vec{p}_{D^-}.\vec{p}_{D_s^+}f^1_{D^-\pi^+,\bar{D}^*(2007)^0}f^1_{D_s^+\bar{D}^*(2007)^0,B^+}}{E_{B^+}-E_{D_s^+}-E_{\bar{D}^*(2007)^0}+\frac{i}{2}\Gamma_{\bar{D}^*(2007)^0}}\\
A_{\bar{D}^*_1(2600)} =& c_{\bar{D}^*_1(2600)} \\
& \times \frac{\vec{p}_{D^-}\cdot \vec{p}_{D_s^+}f^1_{D^+\pi^+,\bar{D}^*_1(2600)}f^1_{D_s^+\bar{D}^*_1(2600),B^+}}{E_{B^+}-E_{D_s^+}-E_{\bar{D}^*_1(2600)}+\frac{i}{2}{\tilde{\Gamma}_{\bar{D}^*_1(2600)}}},
\end{aligned}
\end{equation}
where both $c_{\bar{D}^*(2007)^0}$ and $c_{\bar{D}^*_1(2600)}$ are complex coupling constants that need to be fitted. Here we adopted the experimental energy-dependent width $\tilde{\Gamma}_{\bar{D}^*_1(2600)}$  and will discuss it later. 
Since the width of $\bar{D}^*(2007)^0$ is too small, we used the experimental value.
The $f_{ij}^L$ and $f^L_{ij,k}$ is the dipole form factors defined by:
\begin{equation}\label{eq2}
\begin{aligned}
		f_{ij}^L&=\frac{1}{\sqrt{E_i E_j}}\left( \frac{\Lambda^2}{\Lambda^2+q_{ij}^2} \right)^{2+L/2}\\
f_{ij,k}^L&\equiv f_{ij}^L/\sqrt{E_k}\\
F_{mn}^L & = \frac{1}{\sqrt{E_m E_n}}\left( \frac{\Lambda^2}{\Lambda^2+\tilde{p}_{m}^2} \right)^{2+L/2}
\end{aligned}
\end{equation}
where $q_{ij}(\tilde{p}_m)$ is the momentum of $i(m)$ in the $ij$(total) center-of-mass (COM) frame;
$L$ is the orbital angular momentum of the $ij$ pair system. In calculation, we take a cut-off for $\Lambda$ and set it to common value of $1\GeV$.

As the $\bar{D}^*_2(2460)$ resonance state with the $J^P = 2^+$ , the $B^+ \to D_s^+ \bar{D}^*_2(2460)$ vertex  should be parity-conserving with a $d\text{-wave}$ which amplitude is given by:
\begin{equation}\label{eq3}
\begin{aligned}
A_{\bar{D}^*_2(2460)} =& c_{\bar{D}^*_2(2460)} \frac{\vec{p}^i_{D^-}\vec{p}^j_{D^-}\vec{\epsilon}^{ij}_{\bar{D}^*_2(2460)}\vec{p}^m_{D_s^+} \vec{p}^{n}_{D_s^+} \vec{\varepsilon}^{mn}_{\bar{D}^*_2(2460)}}{E_B-E_{D_s^+}-E_{\bar{D}^*_2}+\frac{i}{2} \tilde{\Gamma}_{\bar{D}^*_2(2460)}}\\
&\times f^2_{D^-\pi^+,\bar{D}^*_2}f^2_{D_s^+\bar{D}^*_2,B^+}.
\end{aligned}
\end{equation}

Summing the spin of the polarization vector~\cite{Molina:2008jw} of the tensor $(2^+)$ meson takes non-relativistic approximation:
\begin{equation}
	\begin{aligned}
		&\sum_{pol} \epsilon_{\mu\nu}\epsilon^*_{\mu'\nu'}\\
		=&\frac{1}{2}\left(\tilde{\delta}_{\mu \mu^{\prime}} \tilde{\delta}_{\nu \nu^{\prime}}+\tilde{\delta}_{\mu \nu^{\prime}} \tilde{\delta}_{\nu \mu^{\prime}}\right)-\frac{1}{3} \tilde{\delta}_{\mu \nu} \tilde{\delta}_{\mu^{\prime} \nu^{\prime}},
	\end{aligned}
\end{equation}
then the amplitude of Eq.~(\ref{eq3}) becames:
\begin{equation}\label{eq4}
	\begin{aligned}
		A_{\bar{D}^*_2(2460)} =& c_{\bar{D}^*_2(2460)} \frac{ (\vec{p}_D \cdot \vec p_{D_s})^2-\frac{1}{3}(\vec{p}_D^2 \vec p_{D_s}^2)}{E_B-E_{D_s^+}-E_{\bar{D}^*_2}+\frac{i}{2} \tilde{\Gamma}_{\bar{D}^*_2(2460)}}\\
		&\times f^2_{D^-\pi^+,\bar{D}^*_2}f^2_{D_s^+\bar{D}^*_2,B^+}.
	\end{aligned}
\end{equation}

We use the mass-dependent running width [see details in~\cite{LHCb:2022lzp}]:
\begin{align}
	\tilde{\Gamma}(M) &= \Gamma_R\left(\frac{q(M)}{q_0}\right)^{2L'+1} \frac{m_R}{M}{F'}^2(M,L'),
\end{align}
where $m_R$ and $\Gamma_R$ are the mass and width of the resonance state, respectively. $M$ is the invariant mass of $D^-\pi^+$ and we denote the orbital angular momentum between $D^-\pi^+$ system and $D_s^+$ by $L'$. The Blatt-Weisskopf form factor~\cite{VonHippel:1972fg} $F'$ takes:
\begin{equation}
	\begin{aligned}
		&F'(M,1) = \sqrt{\frac{1+z^2(M)}{1+z_0^2}} & L=1\\
		&F'(M,2) = \sqrt{\frac{9+3z^2(M)+z^4(M)}{9+3z_0^2+z_0^4}}& L=2,
	\end{aligned}
\end{equation}
where $z(M)=pd$, $z_0=p_0d$, $d$ takes the value used in the experiment of $3.0$ $\text{GeV}^{-1}$~\cite{LHCb:2022lzp}, $p$ is the momentum of particle $D^-$ in the $M_{D^-\pi^+}$ frame, and $q$ denotes the momentum of resonance state in the initial rest frame. The energy dependence widths of $\bar{D}^*_1(2600)$ and $\bar{D}^*_2(2460)$ are as follows:
\begin{equation}
	\begin{aligned}
		&\tilde{\Gamma}_{\bar{D}^*_1(2600)}\\
		=&\Gamma_{\bar{D}^*(2600)} \left(\frac{q_{\bar{D}^*(2600)}}{q_0}\right)^{3} \frac{m_{\bar{D}^*(2600)}}{M_{D^-\pi^+}} F'^2(M_{D^-\pi^+},1)\\
		&\tilde{\Gamma}_{\bar{D}^*_2(2460))}\\
		=&\Gamma_{\bar{D}^*_2(2460)} \left(\frac{q_{\bar{D}^*_2(2600)}}{q_0}\right)^{5} \frac{m_{\bar{D}^*_2(2600)}}{M_{D^-\pi^+}} F'^2(M_{D^-\pi^+},2).
	\end{aligned}
\end{equation}

\subsection{The mechanism of $T^{a}_{c\bar{s}0}(2900)^{{++}}$ amplitude}

We consider the exotic state candidates $T_{c\bar{s}^0(2900)}$ discovered in the experiment as triangle singularities. The amplitude of the triangle diagram in FIG.~\ref{fig1}(b) for $B^+ \to D_s^+D^-\pi^+$ process is given by:
\begin{equation}
	\begin{aligned}
		&A_{D^-D_s^+\pi^+;B^+}\\
		=&c_{T^a_{c\bar{s}0(2900)^+}}\int d\vec{p}_{\chi_{c1}} \frac{v(D_s^+\pi;K^{*+}D^{*+})v(D^{*+}D^-;\chi_{c1})}{E-E(K^{*+})-E(D^{*+})-E(D^-)+\frac{i}{2}\Gamma_{K^{*+}}}\\
		&\times \frac{v(\chi_{c1}K^{*+};B^+)}{E-E(\chi_{c1})-E(K^{*+})+\frac{i}{2}\Gamma_{K^{*+}}+\frac{i}{2}\Gamma_{\chi_{c1}}},
	\end{aligned}
\end{equation}
where the implicit summation of the spin of the intermediate particles are involved. $E$ is the total energy in the COM frame and the energy $E_a(\vec{p}_a)$ is $\sqrt{\vec{p}_a^2+m^2_a}$. $\vec{p}_{\chi_{c1}}$ is a loop momentum. The mass and width values of particles are taken from Particle Data Group (PDG)~\cite{Workman:2022ynf}. 

We use an $s\text{-wave}$ interaction for $D^{*+}(1^-)K^{*+}(1^-) \to D_s^+(0^-)\pi^+(0^-)$ vertex:
\begin{equation}\label{eq10}
	\begin{aligned}
		v(D_s^+\pi^+;K^{*+}D^{*+}) & =\vec \epsilon_{D^{*+}} \cdot \vec \epsilon_{K^{*+}}J_1 f^0_{D_s^+\pi^+}J_2 f^0_{D^{*+}K^{*+}}
	\end{aligned}.
\end{equation}

The vertices of $\chi_{c1}(1^+) \to D^-(0^-)D^{*+}(1^-)(s\text{-wave})$ and  $B^+(0^-) \to \chi_{c1}(1^+) K^{*+}(1^-)(s\text{-wave})$ processes are given as:

	\begin{align}
		\label{eq11}
		v(D^{*+}D^-;\chi_{c1}) &= \vec \epsilon_{\chi_{c1}} \cdot \vec \epsilon_{D^{*+}}^* J_3 f^0_{D^-D^{*+};\chi_{c1}}\\
		\label{eq12}
		v(\chi_{c1}K^{*+};B^+) &= \vec \epsilon_{\chi_{c1}}^* \cdot \vec \epsilon_{K^{*+}}^* f^0_{\chi_{c1}K^{*+};B^+}.
	\end{align}

We use the $M \simeq 4684 \text{MeV}, \Gamma \simeq 126\text{MeV}$ for  $\chi_{c1}(4685)$ state.  

We calculate the interaction vertices of Eq.~(\ref{eq10}) and Eq.~(\ref{eq11}) under the two-body COM frame, and then multiply by a kinematical factors of $J_1, J_2$ and $J_3$ to account for the Lorentz transformation to the COM frame of three-body system~\cite{,Keister:1991sb,Wu:2013xma}. Further details can be found in Ref.~\cite{Kamano:2011ih}.

\subsection{The mechanism of coupled-channel $D\pi$ $S\text{-wave}$ amplitude}

For the $D\pi \ S\text{-wave}$, experimental and theoretical results show that  it cannot be described as a resonance state using a simple BW amplitude, so in this work we consider it as coupled channel effect which contains three channels: $D\pi - \bar{D}_sK -\bar{D}\eta$ in FIG.~\ref{fig1}(c). 
We denote a meson(M)-meson(M) pair with $J^P$ by MM($J^P$), such that $D\pi(0^+)$ denotes a $D\pi$ pair with $J^P= 0^+$. The initial weak vertex $B^+(0^-) \to  D\pi(0^+)D_s^+(0^-) (s\text{-wave})$ is given as:
\begin{equation}\label{eq13}
	\begin{aligned}
		\nu_1 = c^{0^+}_{D\pi D_s^+,B^+} \la t_D \ t_D^z \ t_{\pi} \ t^z_{\pi} | \frac{1}{2} \ \frac{1}{2}\ra f^0_{D\pi} F^0_{D_s^+B^+},
	\end{aligned}
\end{equation}
where $c^{0^+}_{D\pi D_s^+,B^+}$ is a complex coupling constant, $f^0_{D\pi}$ and  $F^0_{D_s^+B^+}$ are form factors presented in Eq.~(\ref{eq2}). An isospin Clebsch-Gordan coefficient is given by the bracket $ \la t_a \ t_a^z \ t_{b} \ t^z_{b} | c \ d\ra$ where $t_a$ and $t_a^z$ are the isospin and $z\text{-component}$ of particle $a$, respectively.

We adopted the method in Ref.~\cite{Nakamura:2022jpd} to describe the hadron scattering process, which is also consistent with the principle of coupled-channel unitarity. We describe hadron interactions in a form that is not constrained by a particular model where all coupling constants are determined from experimental data.

The $s\text{-wave}$ meson-meson interaction potential is given as:
\begin{equation}
	\begin{aligned}
		\nu_{\beta,\alpha} =&   \la t_{\beta_1} \ t^z_{\beta_1} \ t_{\beta_2} \ t^z_{\beta_2} | \frac{1}{2} \ \frac{1}{2} \ra\la t_{\alpha_1} \ t^z_{\alpha_1} \ t_{\alpha_2} \ t^z_{\alpha_2} | \frac{1}{2} \ \frac{1}{2} \ra \\ & \times f^0_{\beta}\left(p'\right)h_{\beta,\alpha} f^0_{\alpha}\left(p\right)  ,
\end{aligned}
\end{equation}
where $\alpha$ and $\beta$ is represent three interaction channels $D\pi$, $\bar{D}_sK$ and $\bar{D}\eta$, $\alpha_1$ and $\alpha_2$ represent the two mesons in channel $\alpha$. $h_{\beta, \alpha}$ represents the coupling constant between $\alpha$ and $\beta$ channels. We describe the coupled-channel effect in terms of $\left[G^{-1}\left(E\right)\right] _{\beta\alpha} = \delta_{\beta\alpha} - h_{\beta,\alpha}\sigma_{\alpha}\left(E\right)$, where
\begin{equation}
	\begin{aligned}
		\sigma_{\alpha}(E) = \sum\limits_{t^z} \int dq q^2 \frac{\la t_{\alpha_1} \ t^z_{\alpha_1} \ t_{\alpha_2} \ t^z_{\alpha_2} | \frac{1}{2} \ \frac{1}{2} \ra ^2 \left[f^0_{\alpha}(q)\right]^2}{E-E_{\alpha_1}\left(q\right) - E_{\alpha_2}\left(q\right)+i\epsilon},
	\end{aligned}
\end{equation} 
where $\sum_{t^z}$ denotes summation of channels with different masses for $D^-\pi^+$ and $\bar{D}^0\pi^0$ charge conjugate. The vertex $D\pi$, $\bar{D}_sK$ and $\bar{D}\eta \to D^-\pi^+$ is:
\begin{equation}\label{eq16}
	\begin{aligned}
		\nu_2 =  h_{D^-\pi^+,\alpha} \la t_{\alpha_1} \ t^z_{\alpha_1} \ t_{\alpha_2} \ t^z_{\alpha_2} | \frac{1}{2} \ \frac{1}{2} \ra f^0_{D\pi}f^0_{\alpha}  .
	\end{aligned}
\end{equation}

The amplitude for diagram of FIG.~\ref{fig1}(c) is:
\begin{equation}\label{eq16}
	\begin{aligned}
		A_{D\pi} =& 4\pi \sum^{D\pi,\bar{D}_sK,\bar{D}\eta}_{\alpha,\beta} h_{D^-\pi^+,\beta} c^{0^+}_{\alpha D_s^+,B^+} f^0_{D^-\pi^+} \left(p_{D^-}\right)\\
		& \times \sigma_{\beta}\left(M_{D^-\pi^+}\right) G_{\beta\alpha}\left(M_{D^-\pi^+}\right)F^0_{D_s^+B^+}.
	\end{aligned}
\end{equation}

\subsection{Invariant msss distributions of the $B^+ \to D_s^+ D^-\pi^+$ decay}

With the amplitudes of the processes we considered above, we can get the total decay amplitude of $B^+ \to D_s^+ D^-\pi^+$ as below:
\begin{equation}\label{eq17}
	\begin{aligned}
		\mathcal{M} =& c_b+\mathcal{M}_{\bar{D}^*(2007)^0} + \mathcal{M}_{\bar{D}^*_1(2600)} +\mathcal{M}_{\bar{D}^*_2(2460)}\\
		&+\mathcal{M}_{T^a_{c\bar{s}0}(2900)}
		 +\mathcal{M}_{D\pi} 
	\end{aligned},
\end{equation}
where $c_b$ is the complex coupling constant representing the contribution of the background in the experiment.

We use the following equation to calculating three-body differential decay width:
\begin{equation}
	\begin{aligned}
	d^2\Gamma_{M}=\frac{1}{(2\pi)^3}\frac{1}{32 E^3}|\mathcal{M}|^2 d^2M_{ab}d^2M_{bc}.
	\end{aligned}
\end{equation}

For a given value of $m^2_{ab}$, the range of $m^2_{bc}$ is determined by:
\begin{equation}
	\begin{aligned}
		&(m_{bc}^2)_{\rm min}=(E_b^*+E_c^*)^2-\left( \sqrt{E_b^{*2}-m_b^2}+\sqrt{E_c^{*2}-m_c^2} \right)^2
		\\
		&(m_{bc}^2)_{\rm max}=(E_b^*+E_c^*)^2-\left( \sqrt{E_b^{*2}-m_b^2}-\sqrt{E_c^{*2}-m_c^2} \right)^2,
	\end{aligned}
\end{equation}
where $E_b^*$ and $E_c^*$ are the energys of particles $b$ and $c$ in the COM frame of the $ab$ pair, respectively.

\section{RESULTS}
\label{III}

\begin{table}
	\renewcommand{\arraystretch}{1.6}
	\tabcolsep=3.mm
	\caption{\label{table1}
		Parameter values for $B^+\to D^-D_s^{+}\pi^+$ models.
	}
	\begin{tabular}{lc}
		Parameters & Values \\
		$c_{\bar{D}^*(2007)^0}$ & $0.84-3.41\,i$\\
		$c_{\bar{D}^*_1(2600)}$ & $-0.62+1.15\,i$\\
		$c_{\bar{D}^*_2(2460)}$ & $(0.18\times 10^{-5}-0.87\times10^{-5}\,i) \GeV^{-2}$ \\
		$c_{T^a_{c\bar{s}0}(2900)^{++}} $ & $(-9.83-4.72\,i) \GeV^2$\\
		$c_b$ & $66.03-27.74\,i$\\
		$c^{0^+}_{D\pi D_s^+,B^+}$ & $42.91-0.67\,i$\\
		$h_{D^-\pi^+,\beta(D\pi)} $ & $0.989$  \\
		$h_{D^-\pi^+,\beta(\bar{D}_sK)}$ & $1.010$\\
	$h_{D^-\pi^+,\beta (\bar{D}\eta)}$ & $0.98$ \\ 
		$ c^{0^+}_{\bar{D}_sK D_s^+,B^+}$ & $-25.45+0.92\,i$\\
		$ c^{0^+}_{\bar{D}\eta D_s^+,B^+}$ & $-28.80 +0.44\,i$\\
		$h_{D\pi,D\pi}$ & $36.99$\\
		$h_{D\pi,\bar{D}_s K}$& $-16.21$\\
		$h_{D\pi,\bar{D}\eta}$ & $-19.55$\\
		$h_{\bar{D}_s K,\bar{D}_s K}$ & $6.35$\\
		$h_{\bar{D}_s K,\bar{D}\eta}$ & $21.60$\\
		$h_{\bar{D}\eta,\bar{D}\eta}$ & $11.93$\\
		$\Lambda$ & $1.00 \GeV$ (fixed) 
	\end{tabular}
\end{table}

\begin{figure*}[htpb]
	\begin{center}
		\includegraphics[width=0.45\textwidth]{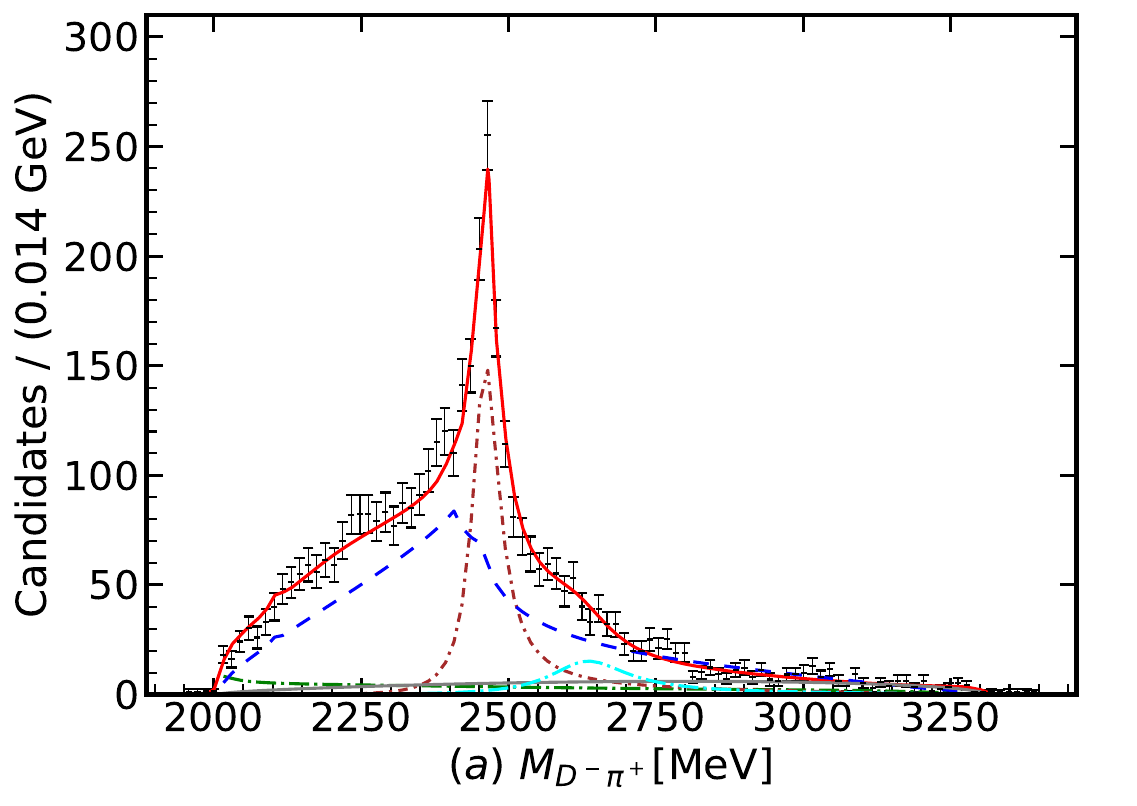}
		\includegraphics[width=0.45\textwidth]{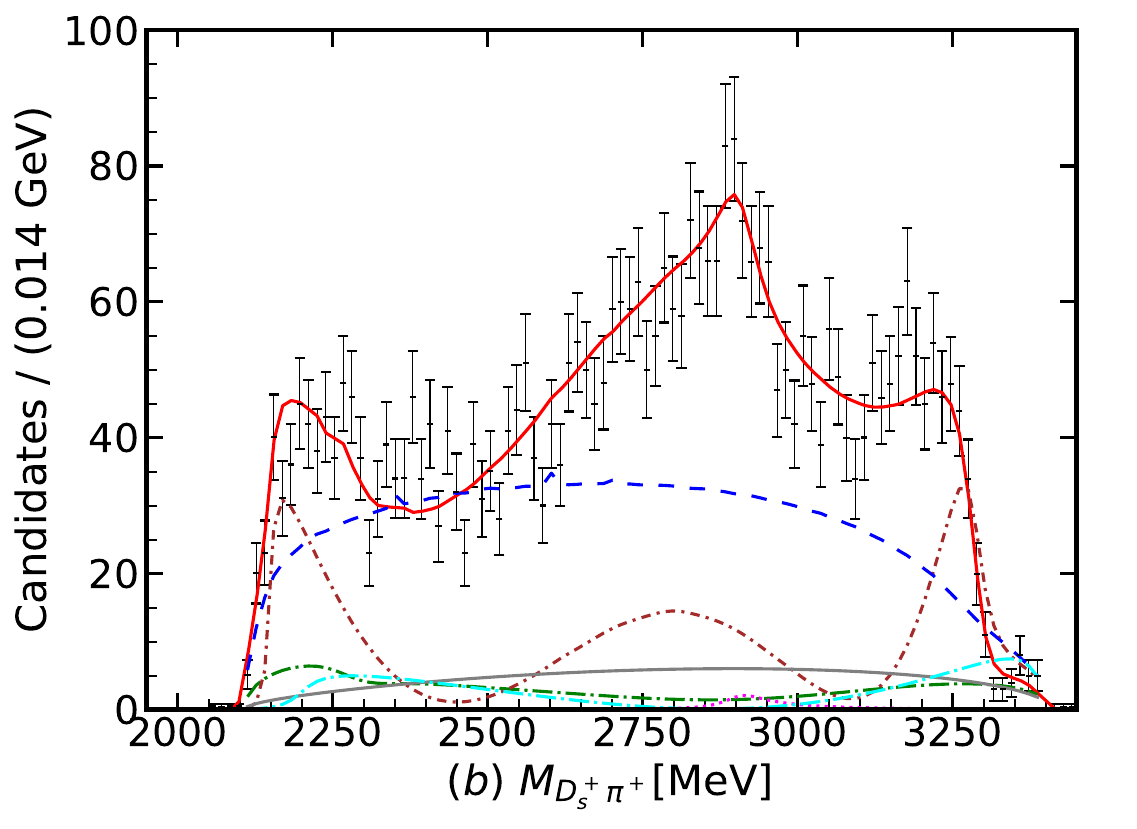}
	\end{center}
	\caption{Comparison with the LHCb data~\cite{LHCb:2022lzp} for $B^+ \to D^-D_s^+ \pi^+$; $(a)\ D^-\pi^+, (b) \ D_s^+\pi^+$ and $(c) \  D^-D_s^+$ invariant mass distributions. The red soild curve is from the full model, smeared with the bin width. Contributions from FIG.~\ref{fig1}$(a)$ are $\bar{D}^*(2007)^0$ [green dashdot], $\bar{D}^*_1(2600)$  [aqua dashdot], and $\bar{D}^*_2(2460)$ [brown dashdotted]. Contribution is from FIG.~\ref{fig1}$(b)$ that includes triangle loop of $T^a_{c\bar{s}0}(2900)^{++}$ [magenta dotted]. Contribution is from FIG.~\ref{fig1}$(c)$ that includes $D\pi (s\text{-wave})$ [blue dotted]. Contribution is from background [grey solid].}
	\label{fig2}
\end{figure*}

\begin{figure}[htpb]
	\begin{center}
		\includegraphics[width=0.45\textwidth]{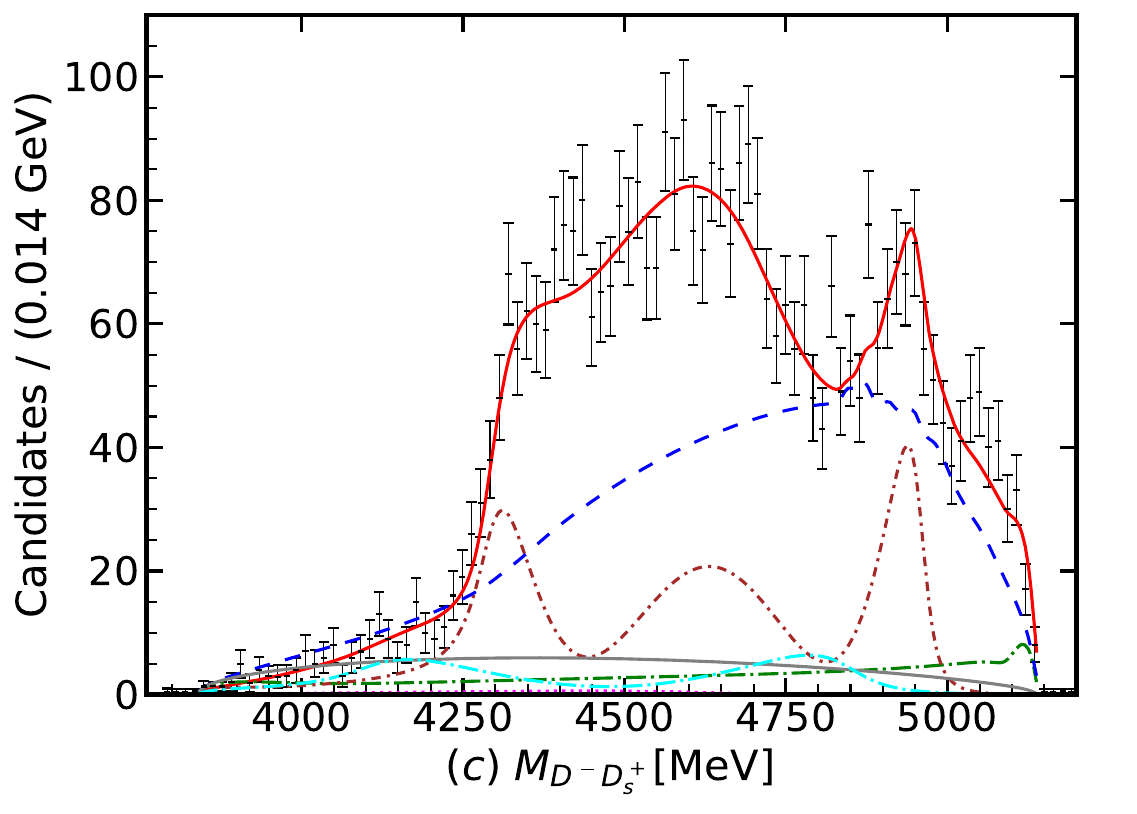}
	\end{center}
	\caption{This figure is a continuation from FIG.~\ref{fig2}.}
	\label{fig23}
\end{figure}

We simlultaneously fit the theoretical model of $B^+\to D^-D_s^{+}\pi^+$ process with the three experimental results of $M_{D^-\pi^+}$, $M_{D_s^+\pi^+}$ and $M_{D^-D_s^+}$ distributions. Theoretical model includes three Breit-Wigner amplitudes, one triangle loop amplitude of $T^{a}_{c\bar{s}0}(2900)^{{++}}$ and one unitary coupled channel effect of $D\pi$ $S\text{-wave}$ amplitude. All amplitudes contain the product of the overall factors of a complex coupling coefficient. In the absence of other experimental inputs, we determine the complex factors by fitting the available data. We take a complex coupling constant($c_b$) for the amplitude in order to represent the background contribution that is considered in the experiment. 
		For different $\beta$ channels, we represent $h_{D^-\pi^+,\beta}$ in Eq.~\ref{eq16} as the three parameters  $h_{D^-\pi^+,\beta(D\pi)} $, $h_{D^-\pi^+,\beta(\bar{D}_sk)}$ and $h_{D^-\pi^+,\beta (\bar{D}\eta)}$ in Table~\ref{table1}.
		In the coupling coefficient of $h_{\beta,\alpha}$ in the $G_{\beta,\alpha}$ term of Eq.~\ref{eq16}, since  $\alpha$ and $\beta$ contain three interaction channels $D\pi$, $\bar{D}_sK$ and $\bar{D}\eta$, there are six parameters $h_{D\pi,D\pi}$, $h_{D\pi,\bar{D}_sK}$, $h_{D\pi,\bar{D}\eta}$, $h_{\bar{D}_sK,\bar{D}_sK}$, $h_{\bar{D}_sK,\bar{D}\eta}$ and $h_{\bar{D}\eta,\bar{D}\eta}$, which are listed in Table~\ref{table1}.
Regarding the $\Lambda$ in the form factors, we adopted a classical value of $1\GeV$. At last, our default model was refined to include a total of $17$ parameters.  Subsequently, we determined the coupling constants from the fit and present numerical results in the \TAB{table1}.

The mass distribution results as shown in FIG.~\ref{fig2} and FIG.~\ref{fig23} are in quite good agreement with the LHCb data, in which the default model is represented by a red soild curves. 
The fit quality is $\chi^2/\text{ndf} = (96.29+111.29+78.01)/(99*3-17) \simeq 1.02$, where three $\chi^2$s are from comparing to the $M_{D^-\pi^+}$, $M_{D_s^+\pi^+}$ and $M_{D^-D_s^+}$ distributions, respectively; 'ndf' represents the number of bins subtracted by the number of fitting parameters. We have taken into account the smearing effect by applying bin widths to theoretical curves for the default model. Overall, the contributions of $\bar{D}^*_2(2460)$ in FIG.~\ref{fig1}(a) [brown dashdotted] and FIG.~\ref{fig1}(c) for $D\pi$ $S\text{-wave}$ [blue dotted] dominate the whole process. 

In the distribution of $M_{D^-\pi^+}$~\ref{fig2}(a), we can clearly note that the contribution of the resonant state $\bar{D}^*_2(2460)$ leads to a sharp peak near $2.46\GeV$. A significant fraction of the amplitude below the energy range of $2.46\GeV$ is attributed to the contributions from the $D\pi$ $S\text{-wave}$ amplitude. In addition, $D^*_1(2600)$ also leads to a comparable peak near $2.6\GeV$. They are indispensable to get a satisfactory fit result.

Then in the $M_{D_s^+\pi^+}$ distribution of $B^+$ decay. The triangular singularity generates a distinct resonance-like peak near the position of $2.9\GeV$ corresponding to $T^{a}_{c\bar{s}0}(2900)^{{++}}$. 
The other contribution of $D^*_1(2600)$ is helpful in improving the cusp in the $3.2\GeV$ region in FIG.~\ref{fig2}(b). The contribution of coupled channel effect of $D\pi$ $S\text{-wave}$ is a large fraction in the whole process. So, for the decay process of fitting $B^+\to D^-D_s^{+}\pi^+$, it is evident that the most crucial aspect is the amplitude contribution of $D\pi$ $S\text{-wave}$. From the fitting results, it appears that our theoretical model provides a highly satisfactory explanation. 

Every $\bar{D}^*_{J}$ resonance state in our analysis is modeled using the BW form and is able to accurately match the experimental data. All known $\bar{D}^{*}_J$ mesons are considered in the experimental model, but the broad $D^*_0(2300)$ state is not included. Because recent experimental studies~\cite{LHCb:2016lxy} and theoretical analyses~\cite{Du:2020pui,Liu:2012zya} suggest that the $D^*_0(2300)$ resonance state is not adequately represented by a simple BW lineshape. For the $D\pi$ $S\text{-wave}$ in the experiment, a quasi-model-independent (qMI) parameterization is employed. This approach divides the $M_{D\pi}$ range into $k$ slices, see Ref.~\cite{LHCb:2016lxy} for more details. We tried to describe the $D\pi$ $S\text{-wave}$ as the BW amplitude of $D^*_0(2300)$ resonance state and the fit value of its simultaneous fit to three invariant mass distrubutions is $\chi^2/\text{ndf} = (207.38+425.27+298.22)/(99 \times 3-14) \simeq 3.29$.  Thus, the BW amplitude is really not a good representation of the $D\pi$ $S\text{-wave}$. 
\\

 \begin{figure}[htpb]	
	\begin{center}
		\includegraphics[width=0.45\textwidth]{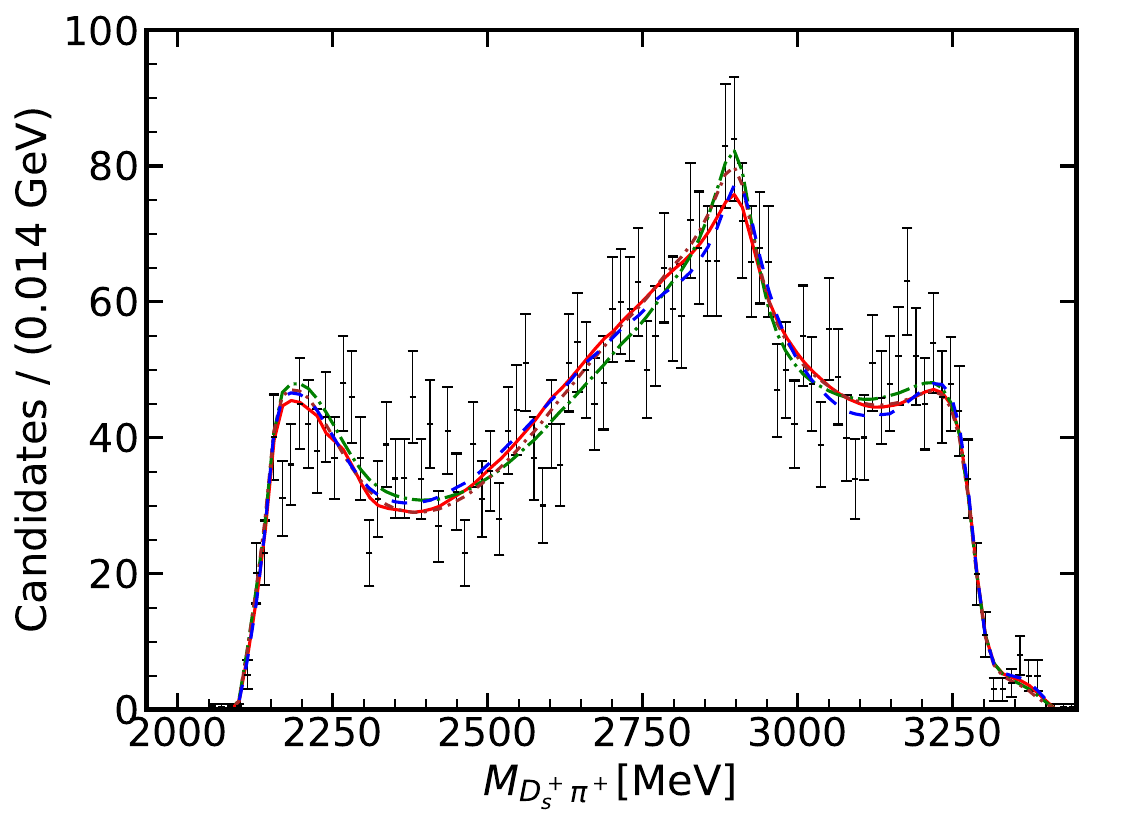}
	\end{center}
	\begin{minipage}{0.45\textwidth} 
		\caption{The $M_{D_s^+\pi^+}$ distribution from the fits
			with different cutoff ($\Lambda$) values in the dipole form factors. The
			blue dotted, red solid, green dotted, and brown dash-dotted
			curves are obtained with $\Lambda=$ 800, 1000,\\ 1250, and 1500 MeV,
			respectively. Other features a-\\re the same as those in FIG. 2(b). }
		\label{fig4}
	\end{minipage}
\end{figure}

 \begin{figure}[htpb]	
	\begin{center}
		\includegraphics[width=0.45\textwidth]{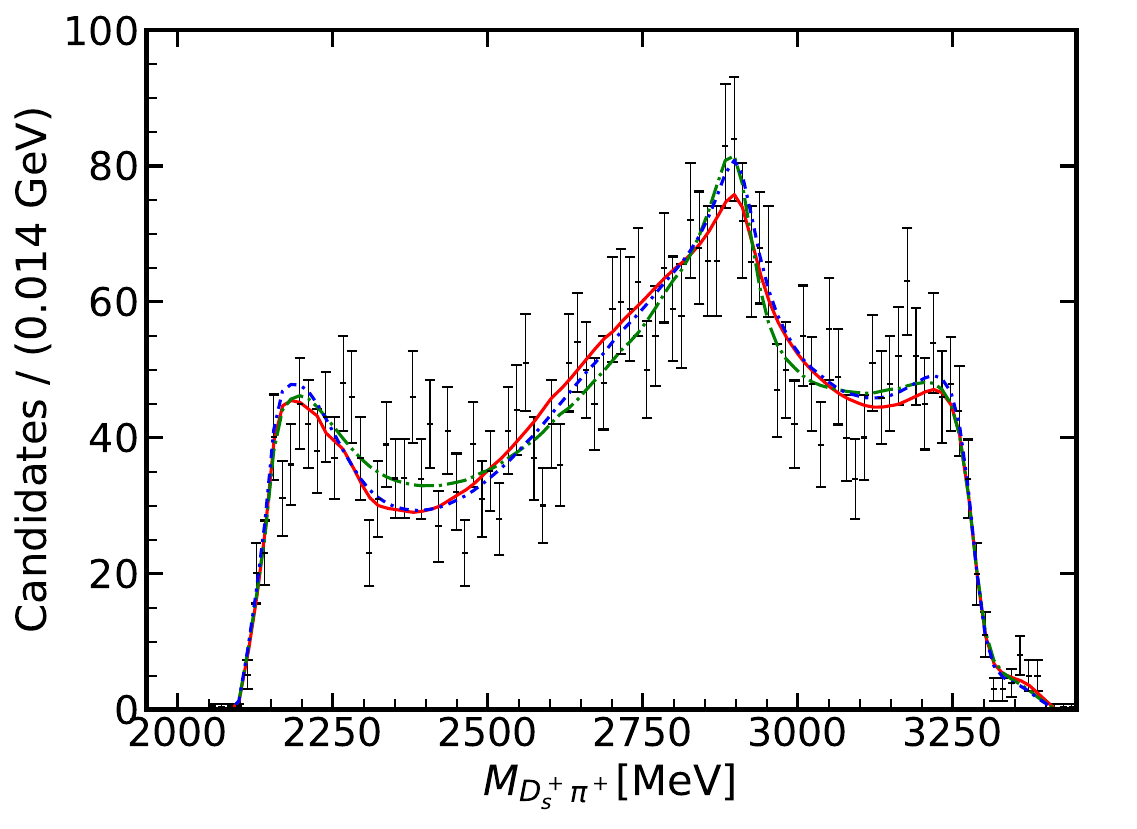}
	\end{center}
	\begin{minipage}{0.45\textwidth} 
		\caption{The $M_{D_s^+\pi^+}$ distribution from the fits
				with different  form
				factors. The
				 red solid, blue dotted, a-\\nd green dash-dotted
				curves are obtained with  dipo-\\le, monopole and Gaussian,
				respectively. Other feat-\\ures are the same as those in FIG. 2(b). }
		\label{fig5}
	\end{minipage}
\end{figure} 

We examine if the fit is stable against changing the form
factor. Instead of $\Lambda=$ 1000 MeV (cutoff) in all the dipole
form factors of the default model, we fit the data with
$\Lambda=$ 800, 1250, and 1500 MeV. As seen in FIG.~\ref{fig4} for the
$M_{D_s^+\pi^+}$ distribution, while the sharpness of the $T(2900)$
peak is somewhat sensitive to the cutoff value, the fit is
reasonably stable overall. Similarly, stable fits are also
obtained for the $M_{D^{-}\pi^+}$ and $M_{D^-D_s^+}$ distributions. 
We also used monopole and Gaussian form factors with
$\Lambda=$ 1000 MeV, and confirmed that the result is very similar to
the case of dipole form factor.  The result is shown in FIG.~\ref{fig5}.

\begin{figure}[htpb]	
	\begin{center}
		\includegraphics[width=0.45\textwidth]{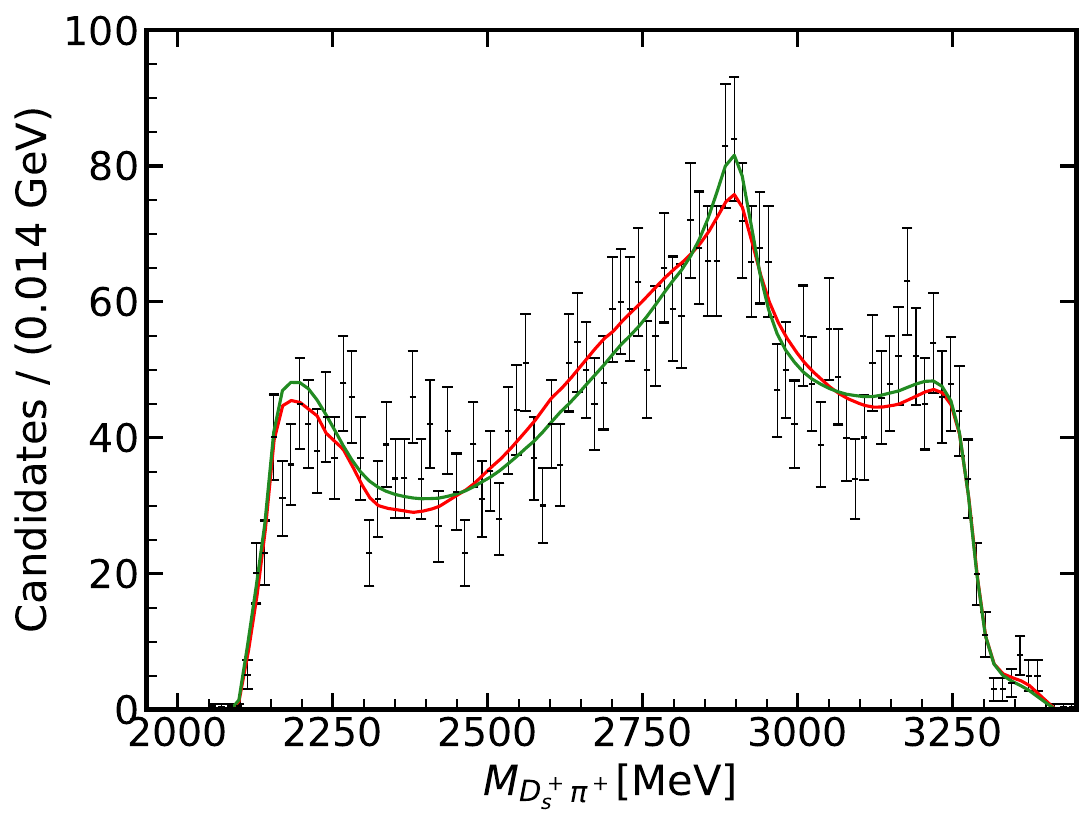}
	\end{center}
	\begin{minipage}{0.45\textwidth} 
		\caption{The figure shows the $M_{D_s^+\pi^+}$ distribution o-\\btained from the combined fit of the different model-\\s, with the red curve representing the default model and the green curve representing the variation of th-\\e $\Lambda$ and $d$ parameters included in the fit over a  reas-\\onable range. }
		\label{fig6}
	\end{minipage}
\end{figure}

  \begin{figure}[htpb]	
	\begin{center}
		\includegraphics[width=0.45\textwidth]{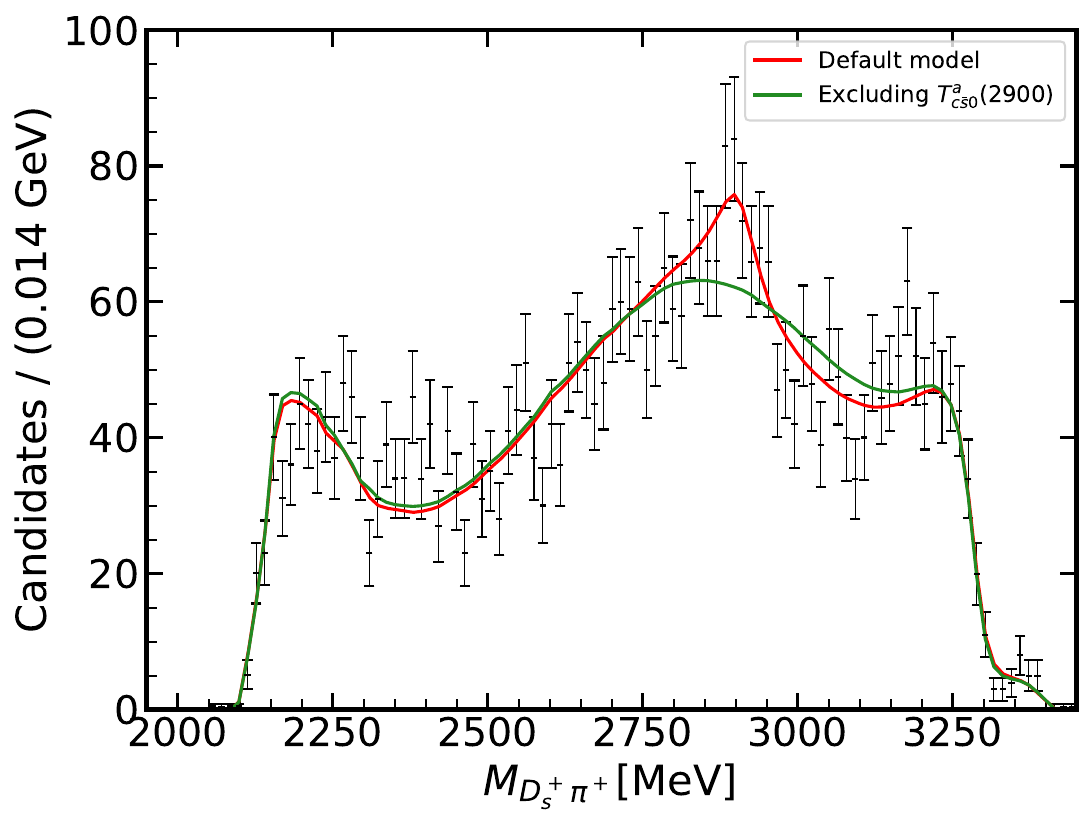}
	\end{center}
	\begin{minipage}{0.45\textwidth} 
		\caption{The $M_{D_s^+\pi^+}$ distribution from combined fits for different models is shown, with the red curve rep-\\resenting the full model and the green curve represe-\\nting the full model with $T^{a}_{c\bar{s}0}(2900)^{{++}}$ contribution removed. }
		\label{fig7}
	\end{minipage}
\end{figure}
In order to have a thorough understanding of the model, we refit the total amplitude after taking   into account the cutoff $\Lambda$ and the parameter $d$ in the fit and the result for the $M_{D_s^+\pi^+}$ distribution is shown in FIG.~\ref{fig6}. We explored the variation of the parameter $d$ in the range 1.5 to 5 GeV and $\Lambda$ in the range 500 to 2000 MeV.
From the results it looks like it will improve the sharp at 2.9 GeV, overall it looks the same as in the default model in FIG.~\ref{fig2}(b).

 In order to test the importance of $T^{a}_{c\bar{s}0}(2900)^{{++}}$, we refit the total amplitude after removing the amplitude of $T^{a}_{c\bar{s}0}(2900)^{{++}}$ and the result  is shown in FIG.~\ref{fig7} that we compare the $M_{D_s^+\pi^+}$ contribution of the default model with the amplitude that does not include $T^{a}_{c\bar{s}0}(2900)^{{++}}$.The fit quality is  $\chi^2/\text{ndf} = (93.23+130.06+80.39)/(99 \times 3-17) \simeq 1.08$. The value of $\chi^2$ in the $M_{D_s^+\pi^+}$ distribution increases from $111.29$ to $130.06$, and it is clear from the FIG.~\ref{fig7} that the fit result becomes significantly worse around $2.9$ $\GeV$ energy region. Despite its minimal fraction in FIG.~\ref{fig2}(b), it still played a role in improving the shape of the peak at the $2.9\GeV$ area.

 \section{CONCLUSION}
 \label{IIII}
 We have made a theoretical study of the $B^+ \to D^-D_s^+\pi^+$ reaction recently researched by the LHCb. 
  The triangular loop mechanism in the model causes a TS peak near the $D^{*+}K^{*+}$ threshold that fits well the peak at $2900\text{MeV}$ in the invariant mass distribution of  $D_s^+\pi^+$ data.
In order to  investigate the consistency of our model with experimental measurements, we investigate the $D^-\pi^+, D_s^+\pi^+$ and $D^-D_s^+$ invariant mass distributions and fit the experimental data using the parameters mentioned in the formula, and found that there is agreement with experiment at the peaks and dips in all three invariant mass distributions. 
We use a unitary coupled-channel model to characterize the main amplitude contribution of the $D\pi$ $S\text{-wave}$, and obtain good fitting results with as few parameters as possible, which also shows that it is reasonable to use the coupled-channel model to describe the amplitude of the $D\pi$ $S\text{-wave}$. 

Moreover, we have compared the default model with another model that eliminates the contribution of $T_{c\bar{s}0}(2900)^{++}$ to confirm the necessity of both amplitudes in our default model. The results indicate that the contribution of $T_{c\bar{s}0}(2900)^{++}$ is crucial.

 \begin{acknowledgments}
 	\noindent
 	 XL is supported by the National Natural Science Foundation of China under Grant No. 12205002.
 	HS is supported by the National Natural Science Foundation of China (Grant No.12075043, No.12147205).
 \end{acknowledgments}

\bibliography{ref}
\end{document}